\documentclass[epj,nopacs,final]{svjour}
\usepackage{graphicx}
\usepackage{color,amsmath}

\title{Double-polarization observable \boldmath$G$
in  neutral-pion photoproduction off the proton}
\titlerunning{Measurement of the double-polarization observable $G$
in  neutral-pion photoproduction}
\authorrunning{A. Thiel et al.}

\author{\mail{thiel@hiskp.uni-bonn.de}%
A.~Thiel\inst{1},
  H.~Eberhardt\inst{2}, 
  M.~Lang\inst{1},
F.~Afzal\inst{1},
 A.V.~Anisovich\mbox{\inst{1}$^,$\inst{3}},
 B.~Bantes\inst{2},
 D.~Bayadilov\mbox{\inst{1}$^,$\inst{3}},
   \mbox{R.~Beck\inst{1},}
M.~Bichow\inst{4},
  K.-T.~Brinkmann\mbox{\inst{1}$^,$\inst{7}},
  S.~B\"ose\inst{1},
   V.~Crede\inst{6},
M.~Dieterle\inst{5},
 H.~Dutz\inst{2},
 D.~Elsner\inst{2},
 R.~Ewald\inst{2},
  K.~Fornet-Ponse\inst{2},
  \mbox{St.~Friedrich\inst{7},}
 F.~Frommberger\inst{2},
  \mbox{Ch.~Funke\inst{1},}
  St.~Goertz\inst{2},
 M.~Gottschall\inst{1},
  \mbox{A.~Gridnev\inst{3},}
M.~Gr\"uner\inst{1},
 E.~Gutz\mbox{\inst{1}$^,$\inst{7}},
D.~Hammann\inst{2},
 Ch.~Hammann\inst{1},
 J.~Hannappel\inst{1},
 J.~Hartmann\inst{1},
  W.~Hillert\inst{2},
 Ph.~Hoffmeister\inst{1},
 Ch.~Honisch\inst{1},
 T.~Jude\inst{2},
 D.~Kaiser\inst{1},
 \mbox{H.~Kalinowsky\inst{1},}
F.~Kalischewski\inst{1},
 S.~Kammer\inst{2},
  I.~Keshelashvili\inst{5},
P.~Klassen\inst{1},
 V.~Kleber\inst{2},
  F.~Klein\inst{2},
 E.~Klempt\inst{1},
K.~Koop\inst{1}, 
  \mbox{B.~Krusche\inst{5},}
M.~Kube\inst{1},
 I.~Lopatin\inst{3},
Ph.~Mahlberg\inst{1},
 \mbox{K.~Makonyi\inst{7},}
  \mbox{V.~Metag\inst{7},}
    \mbox{W.~Meyer\inst{4},}
 J.~M\"uller\inst{1},
J.~M\"ullers\inst{1},
  \mbox{M.~Nanova\inst{7},}
   V.~Nikonov\mbox{\inst{1}$^,$\inst{3}},
  \mbox{D.~Piontek\inst{1},}
  S.~Reeve\inst{2},
  G.~Reicherz\inst{4},
  S.~Runkel\inst{2},
   A.~Sarantsev\mbox{\inst{1}$^,$\inst{3}},
 Ch.~Schmidt\inst{1},
  H.~Schmieden\inst{2},
 T.~Seifen\inst{1},
 V.~Sokhoyan\inst{1},
K.~Spieker\inst{1},
 U.~Thoma\inst{1},
M.~Urban\inst{1},
 H.~van Pee\inst{1},
 D.~Walther\inst{1},
 Ch.~Wendel\inst{1},
A.~Wilson\mbox{\inst{1}$^,$\inst{6}},
 A.~Winnebeck\inst{1}~and
L.~Witthauer\inst{5}.\\
 (The CBELSA/TAPS collaboration)\\
}                     
\institute{\inst{1}Helmholtz-Institut f\"ur Strahlen- und Kernphysik der Universit\"at Bonn, Germany\\
\inst{2}Physikalisches Institut, Universit\"at Bonn, Germany\\
\inst{3}Petersburg Nuclear Physics Institute, Gatchina, Russia\\
\inst{4}Institut f\"ur Experimentalphysik I, Ruhr--Universit\"at Bochum,
Germany\\
\inst{5}Institut f\"ur Physik, Universit\"at Basel, Switzerland\\
\inst{6}Department of Physics, Florida State University, Tallahassee, USA\\
\inst{7}II. Physikalisches Institut, Universit\"at Gie{\ss}en, Germany}

\date{Received: date / Revised version: date}

\abstract{
This paper reports on a measurement of the double-polarization observable $G$ in $\pi^0$ photoproduction off the proton using the CBELSA/TAPS experiment at the ELSA accelerator in Bonn. The observable $G$ is determined from reactions of linearly-polarized photons with longitudinally-polarized protons. The polarized photons are
produced by bremsstrahlung off a properly oriented diamond radiator. A frozen spin butanol target provides the polarized protons. The data cover the photon energy range from 617 to 1325\,MeV and a wide angular range. The experimental results for \textit{G} are compared to predictions by the Bonn-Gatchina (BnGa), J\"ulich-Bonn (J\"uBo), MAID and SAID partial wave analyses. Implications of the new data for the pion photoproduction multipoles are discussed.
}
\begin{document}
\maketitle
%
\section{Introduction}

The spectrum of meson \cite{Klempt:2007cp,Crede:2008vw} and baryon \cite{Klempt:2009pi,Crede:2013kia} 
resonances originates from the dynamics of quarks and gluons in the non-perturbative
regime of QCD. The quark model gives
a good estimate of the number of states to be expected and of their approximate
masses. However, in meson spectroscopy, there is possibly an excess in the number of 
observed states compared to quark model predictions; in baryon 
spectroscopy, the quark model predicts \cite{Capstick:1986bm,Loring:2001kx,Giannini:2015zia} the existence of many more states than have been observed experimentally, a fact known as the problem of the {\it missing resonances}. Recent lattice gauge calculations of the baryon spectrum \cite{Edwards:2011jj} confirm an excess in the number of resonances. However, the pion mass used in the simulations is still rather large. Possibly, the missing resonances have a small coupling to $\pi N$ and may therefore have escaped discovery. Photoproduction may hence be a better method for the search for additional baryon resonances. It is not yet clear if dynamically-generated resonances \cite{Sarkar:2009kx,Oset:2009vf}
\begin{figure}[pt]
\resizebox{0.48\textwidth}{!}{%
  \includegraphics{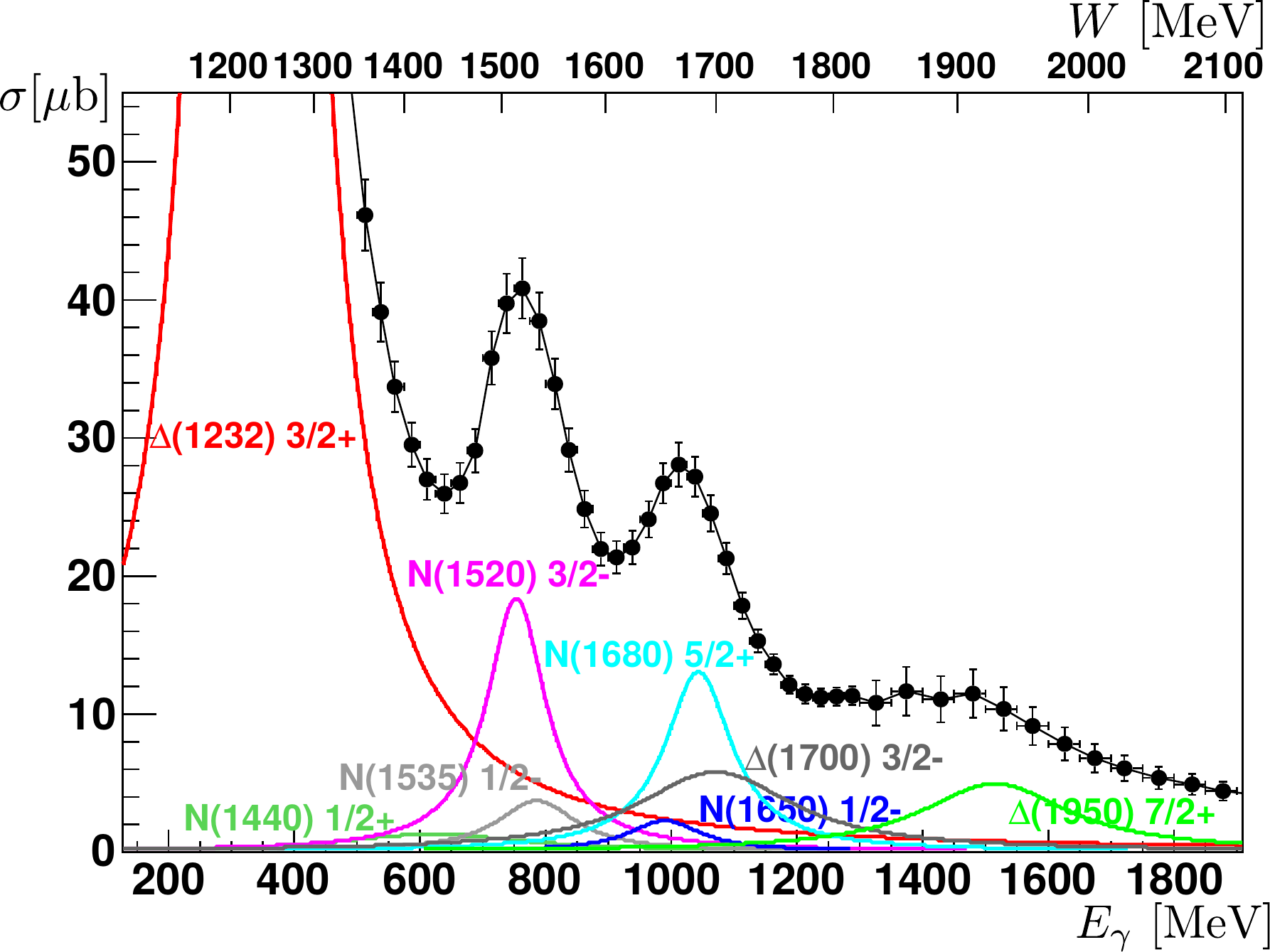}
}
\caption{\label{pic:wq-resonances}The calculated Breit-Wigner distributions of the most important resonances contributing to the total cross section for $\pi^0$ photoproduction off the proton. Data points: \cite{wq_pi0,wq_pi02}, 
resonance parameters from \cite{pdg2014}.} 
\end{figure}%
come on top of these quark model states or whether they are a subclass. Above the mass of about 1.9\,GeV, baryonic hybrids are predicted \cite{Capstick:2002wm}, states in which the gluonic degrees of freedom could manifest themselves in the form of constituent glue. This prediction leads to an even larger number of missing resonances.

The excited states cannot be directly measured since they are broad and strongly overlapping (see Fig.~\ref{pic:wq-resonances}). Data with no polarization information give only access to the quadratic sums of the amplitudes, allowing the determination of the dominant resonant contributions only. It is therefore decisive
to measure many different polarization observables since these are sensitive to interferences of the amplitudes. The large number of observables makes photoproduction particularly well suited to search for weak\-ly coupling resonances. Only when a sufficient number of independent observables is measured with high precision,
the four independent photoproduction amplitudes \cite{Barker:1975bp,Chiang:1996em}
or at least the multipoles (up to a maximum orbital angular momentum) 
\cite{Omelaenko:1981cr,Wunderlich:2014xya}
can be determined without relying on model assumptions.

During the last decade, a new generation of double-polarization experiments, e.g. CLAS at JLab \cite{CLAS}, CB at MAMI \cite{CBMAMI} and CBELSA/TAPS in Bonn \cite{CBELSA}, were built and have accumulated a large amount of data. They partly continue to take further data \cite{CBMAMI,CBELSA}. These experiments are all able to measure reactions with a polarized photon beam and polarized target nucleons (protons or neutrons). Even first experiments with measurements of the recoil polarization of the proton have been carried out \cite{Sikora:2013vfa}. This leads to a new era in baryon spectroscopy since more precise data, covering a large energy and angular range, have become accessible.

The first publications of double-polarization data were mostly focused on the photoproduction of single pseudoscalar mesons, of Kaons~\cite{Zegers:2003ux,Sumihama:2005er,Bradford:2006ba,Lleres:2008em,Ewald:2014ozu}, neutral pions~\cite{thiel:2012,Gottschall:2013uha,hartmann:2014mya,Strauch:2015zob}, or $\eta$-mesons \cite{Akondi:2014ttg,Muller:2015tbd}.

The photoproduction of a single $\pi^0$ was considered to be a well-known reaction, and it was rather surprising that even for this final state substantial disagreements between data and predictions were found. The predictions of different partial wave analyses differed significantly \cite{thiel:2012}. The new polarization data provide strong constraints for the partial wave analyses (PWA) of the different PWA-groups \cite{Hilt:2013coa,Cao:2013psa,Workman:2012jf,Ronchen:2014cna,Kamano:2013iva,Anisovich:2011fc,Anisovich:2013vpa,hartmann:2014b}. This will allow a better determination of the parameters of the resonances contributing to the $p \pi^0$ final state.

In this paper we report on a measurement of the double-polarization variable $G$ for the reaction
\begin{align}
\label{reaction}
\overrightarrow{\gamma} \overrightarrow{p} \rightarrow p \pi^0.
\end{align}
The measurement requires linearly-polarized photons and longitudinally-polarized protons.
A part of the data has been presented in a letter \cite{thiel:2012}. Here, we give a detailed account of the extraction of the beam asymmetry $\Sigma$ and of the double-polarization observable $G$ and present the full data set. 

In an independent paper, results on the double-pola\-ri\-zation variable $E$ will be presented \cite{Gottschall:2015tbd} which uses circularly-polarized photons and longitudinally-polarized protons. The experimental setup and the event reconstruction are fully described there. In section \ref{sec:data}, we give a brief account of the experiment and of the reconstruction of the events due to reaction~(\ref{reaction}). 
Section \ref{sec:extraction_observables} explains how the dilution factor is determined which represents the fraction of events assigned to photoproduction of a single neutral pion off free protons in the target, and describes 
the extraction of the polarization observables, of the beam asymmetry $\Sigma$ and of the double-polarization observable $G$. In section \ref{sec:g}, the results are presented and compared to fits and predictions by different partial wave analyses. The paper concludes with a short summary.

\begin{figure}[th]
\resizebox{0.48\textwidth}{!}{%
  \includegraphics{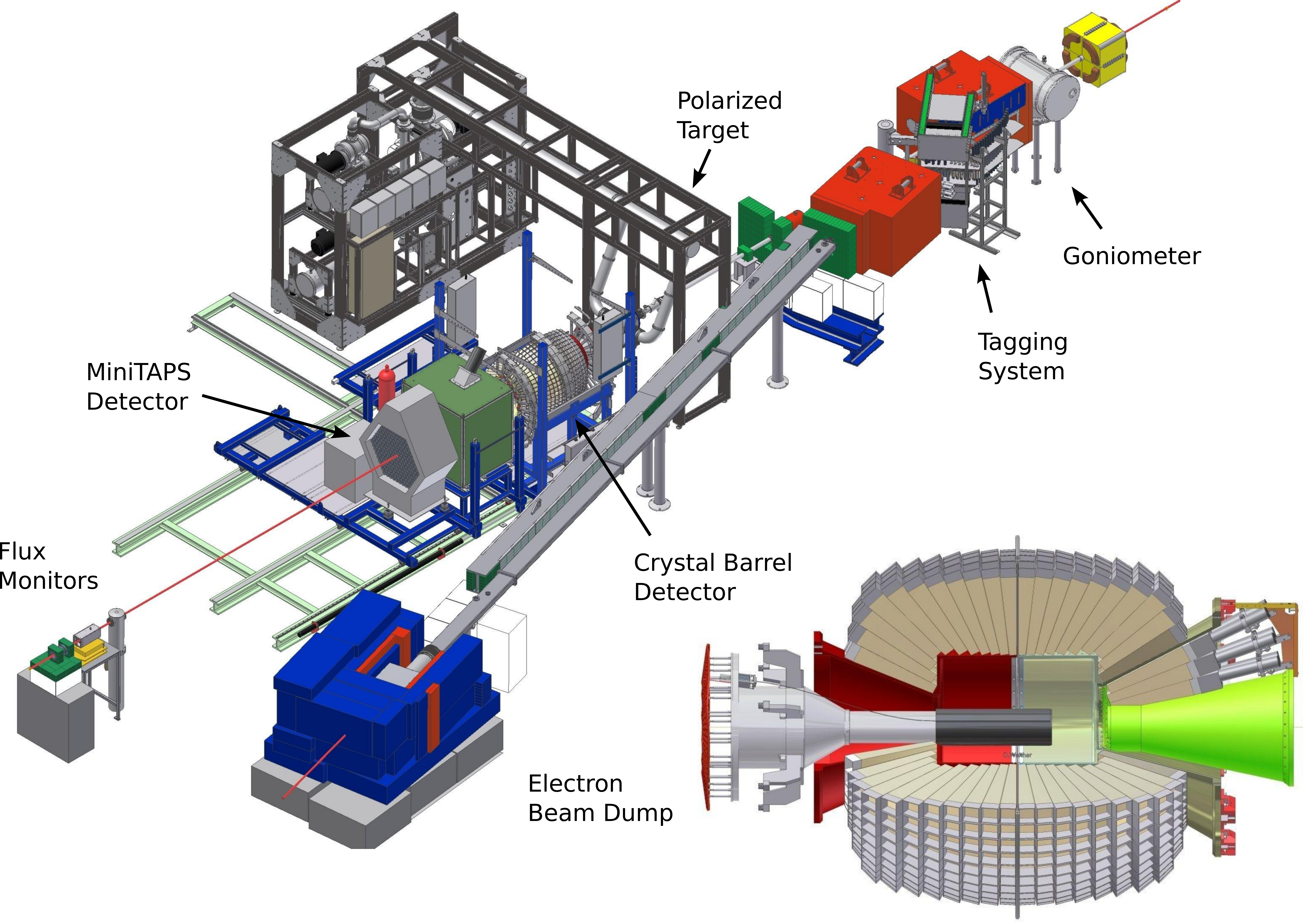}
}
\caption{The setup of the CBELSA/TAPS experiment.}
\label{pic:experimental-setup}
\end{figure}%

\section{\label{sec:data}Experimental Setup and Data Analysis}
\subsection{\label{sec:exp}Experimental Setup}%

The CBELSA/TAPS experiment is shown in Fig. \ref{pic:experimental-setup}. It was located at the ELectron Stretcher Accelerator ELSA in Bonn \cite{Hillert:2006yb} which provided a  $E_0=3.2$\,GeV electron beam. The photon beam was produced by bremsstrahlung off different radiator targets which were mounted on a goniometer wheel. The energy of the generated photons was determined by a tagging hodoscope \cite{KFP:2009} with an energy resolution of better than $0.4\%\cdot E_\gamma$. The goniometer contained horizontal and vertical wires for the beam alignment. 

\begin{figure}[pt]
\resizebox{0.45\textwidth}{!}{%
  \includegraphics{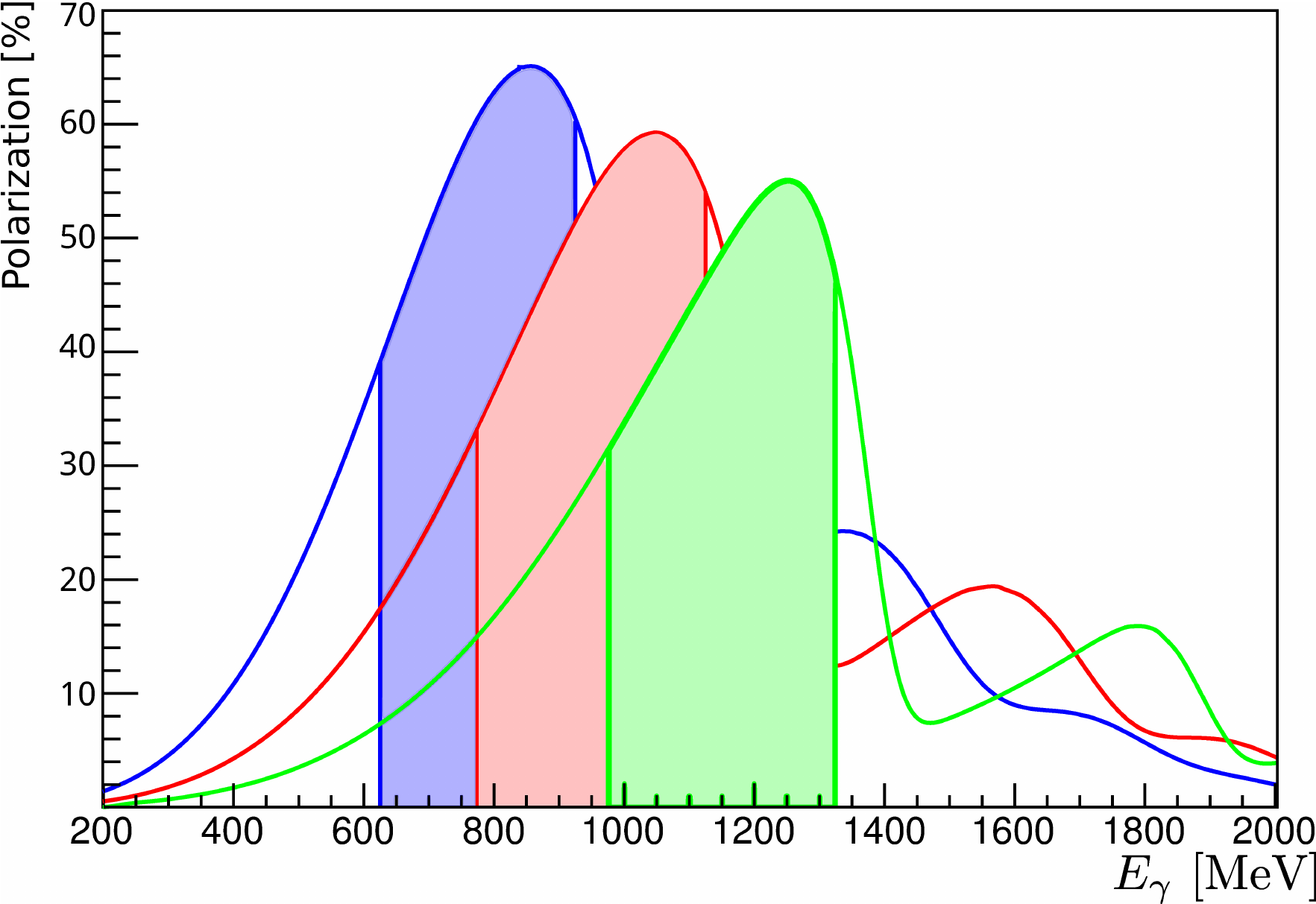}
}
\caption{The three different settings of the coherent edges, the shaded areas mark the energy regions used in this analysis. }
\label{pic:coherentedges}
\end{figure}

Linearly-polarized photons were produced by coherent brems\-strahlung off a diamond crystal~\cite{Elsner:2008sn}. Additional data with an amorphous radiator were taken in order to determine the coherent contributions. The coherent spectrum was extracted by dividing the spectra of the diamond crystal by the amorphous ones. Three different configurations of the coherent edge were used (see Fig.~\ref{pic:coherentedges}): at 950~MeV with a maximal polarization of 65\%, at 1150~MeV with 59\% and at 1350~MeV with 56\%. The degree of polarization was extracted using a software \cite{Elsner:2008sn} based on the analytic bremsstrahlung 
(ANB) calculation \cite{ANB}. To reduce systematic effects, the plane of the linearly-polarized photons was switched by 90$^\circ$ every 15 minutes.

\begin{figure}
\centering
\includegraphics[width=0.45\textwidth]{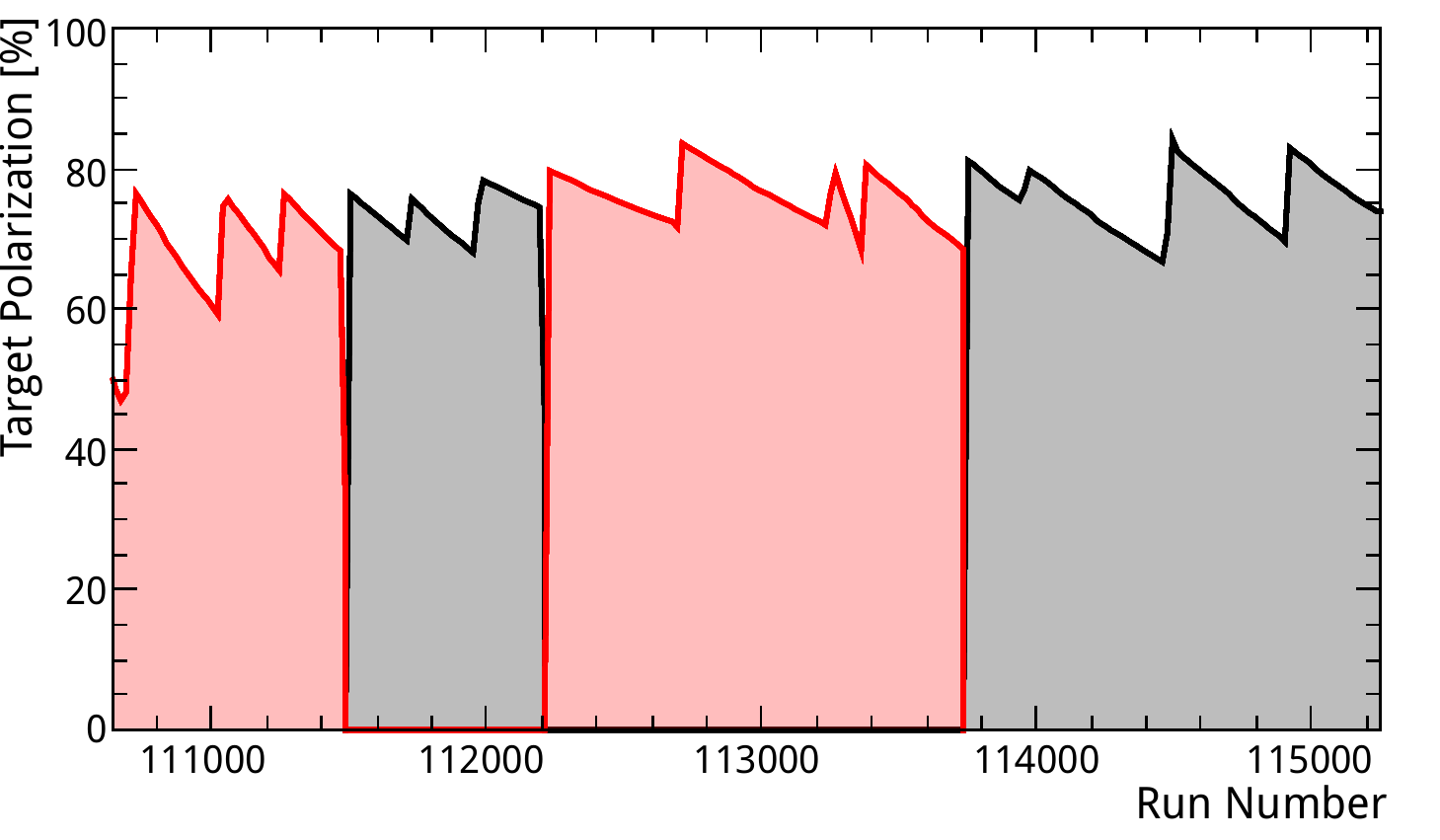}
\caption{The target polarization during one beamtime of 40 days with positive (red) and negative (black) polarization.}\label{pic:targetpol}
\end{figure}%

Beam photons impinged on the longitudinally polarized frozen spin butanol (C$_4$H$_9$OH) target \cite{Bradtke:1999zg}, which was contained within a horizontal $^3$He/$^4$He dilution refrigerator. The protons were dynamically polarized in a high magnetic field typically up to 80\% . The target polarization was
measured by an NMR system with a precision of 1,5\%.
During the measurements the polarization was maintained in the frozen spin mode by an internal superconducting 'holding coil'. Relaxation times up to 600 hours could be reached and the polarization has been refreshed every 48 hours. 
The polarization direction of the protons was switched multiple times during a beam time, to reduce systematic effects in the data. Fig.~\ref{pic:targetpol} shows the degree of target polarization during a data-taking period.
For background studies, a carbon foam target \cite{Gruener2015} was alternatively inserted into the cryostat, which had the same mass area density as the carbon amount in the butanol target.

The target was located at the center of the detector system consisting of 
two calorimeters (Crystal Barrel and MiniTAPS) which covered the full azimuthal range and a polar angular range from 1$^\circ$ to 156$^\circ$ with respect to the photon beam axis. The forward crystals were covered by plastic scintillators in front of each crystal to identify charged particles. Inside the calorimeter, a three-layer inner detector with 513 scintillating fibers was installed. At the end of the photon beam line, a Gamma Intensity Monitor (GIM) measured the photon flux. Additionally, a thin conversion target with multiple scintillating detectors were used, counting a known fraction of the total flux by exploiting Compton scattering and pair production. 

In the data analysis, events were selected with a beam photon in the $617$ to $1325$\,MeV  energy range where linearly-polarized photons were available. For photon energies $617 < E_\gamma < 1117$\,MeV, a $33$\,MeV wide energy binning was chosen, for $1125$\,MeV$ < E_\gamma$, a $50$\,MeV binning was used.

During data-taking, a trigger on events with at least two clusters in the calorimeter system was used. Additionally, a Cherenkov detector between the two calorimeters was used as a veto to reduce background due to pair production.

\subsection{\label{reconstruction}Event reconstruction}
\begin{figure}[htb]
\includegraphics[width=0.4\textwidth]{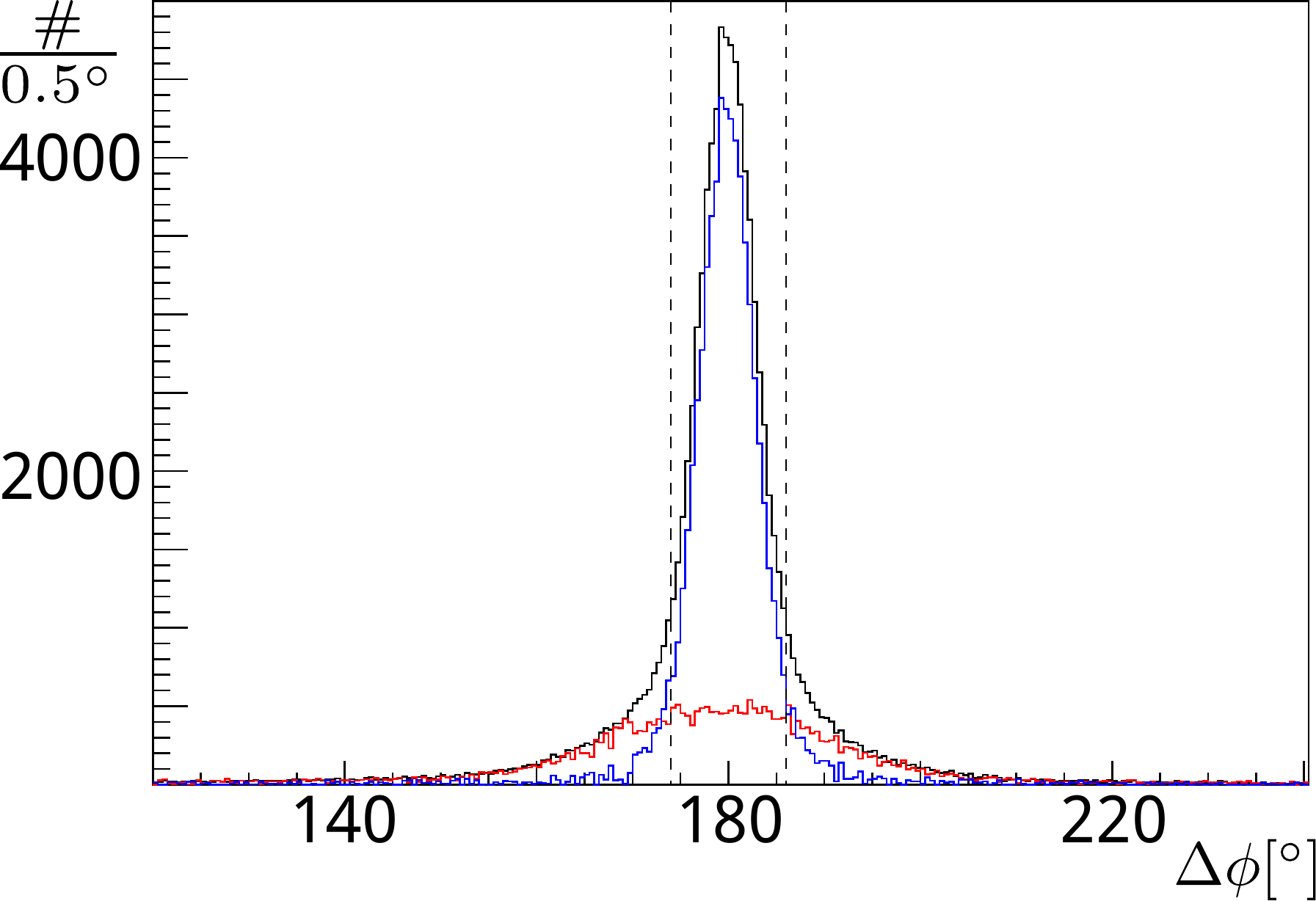}\\[10pt]
\includegraphics[width=0.4\textwidth]{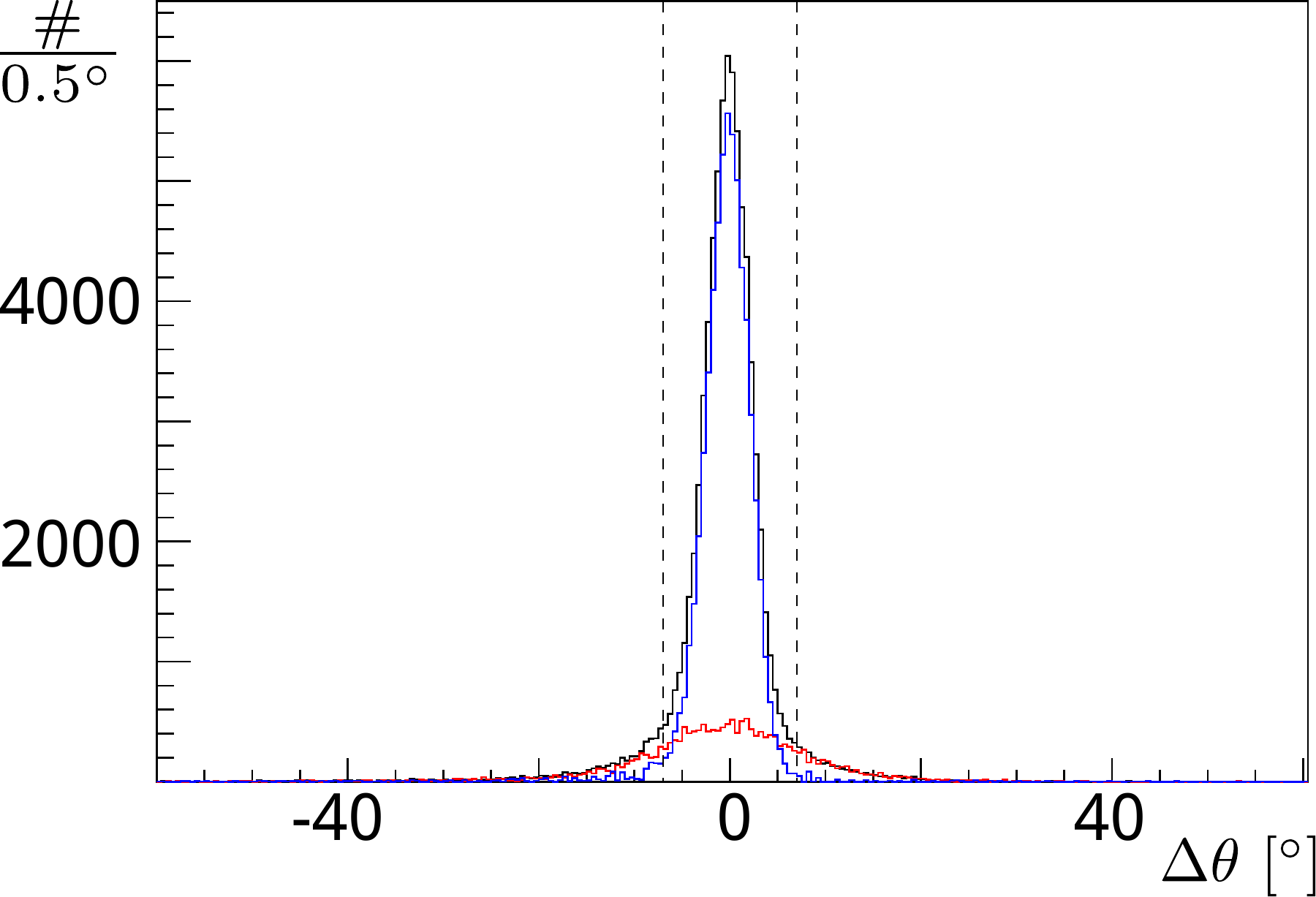}
\caption{\label{pic:kinematicalcuts}The distribution of the directional difference $\Delta\phi$ between the meson and proton momenta (top) and the directional difference $\Delta\theta$ between the measured and the calculated proton momenta (bottom) for the butanol data (black), the carbon data (red) and the difference (blue) for a photon energy of $E_\gamma = 1000 \pm 16$~MeV. \vspace{3mm}}
\end{figure}%

Reaction~(\ref{reaction}) was identified via the $\pi^0 \rightarrow \gamma \gamma$ decay mode. In a first step, events with three or two hits in the calorimeters were selected. Two neutral hits were assigned to the decay photons of the pion. Events with an additional hit in either the calorimeters or the charge sensitive detectors were selected if the third hit had been charged. This hit was assigned to the proton. For events with only two neutral particles in the calorimeters, no further kinematical constraints on the direction of the particles could be applied. These events will later be referred to as two-PED events, all events with a third, charged hit are the so-called three-PED events.

\begin{figure}[tb]
\includegraphics[width=0.4\textwidth]{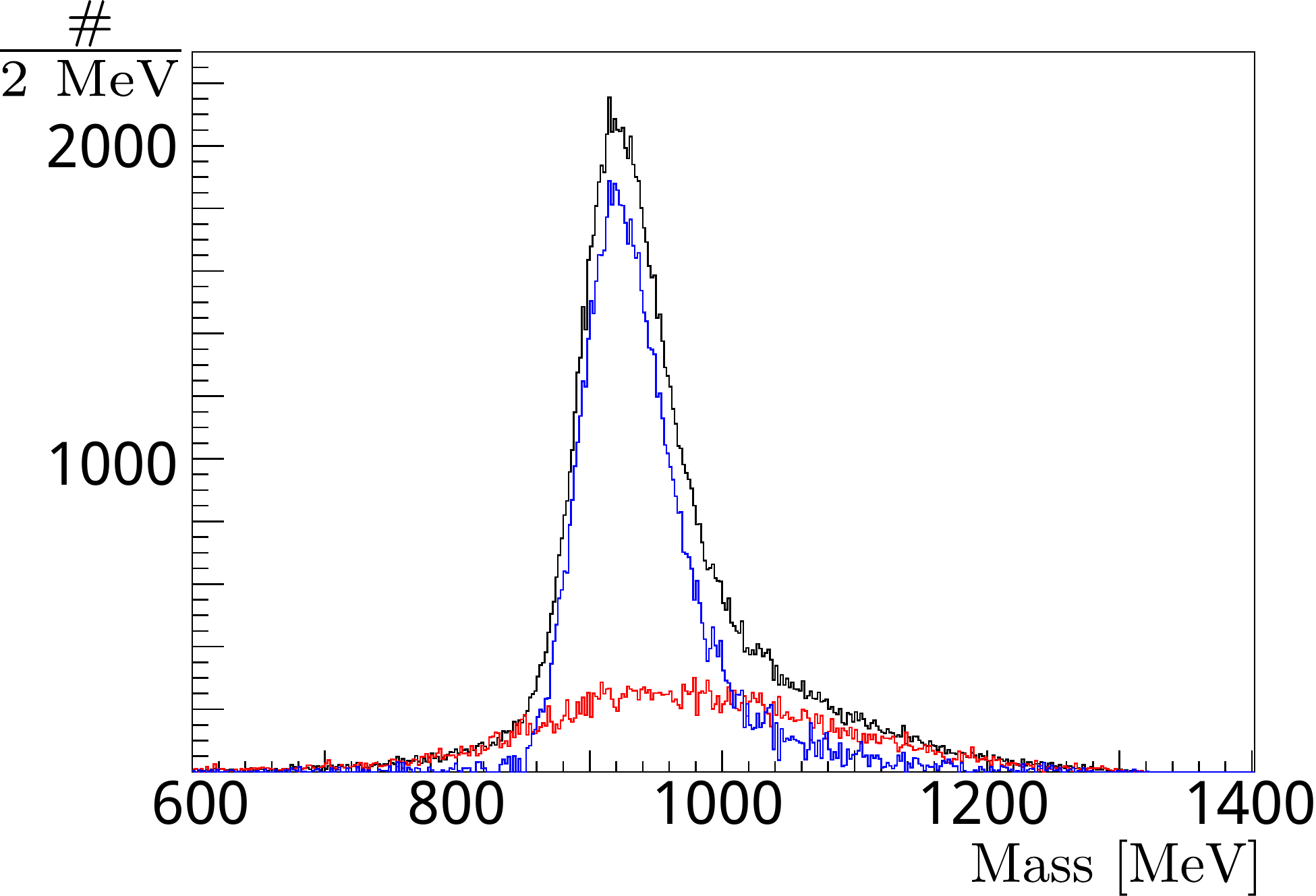} \\
\includegraphics[width=0.4\textwidth]{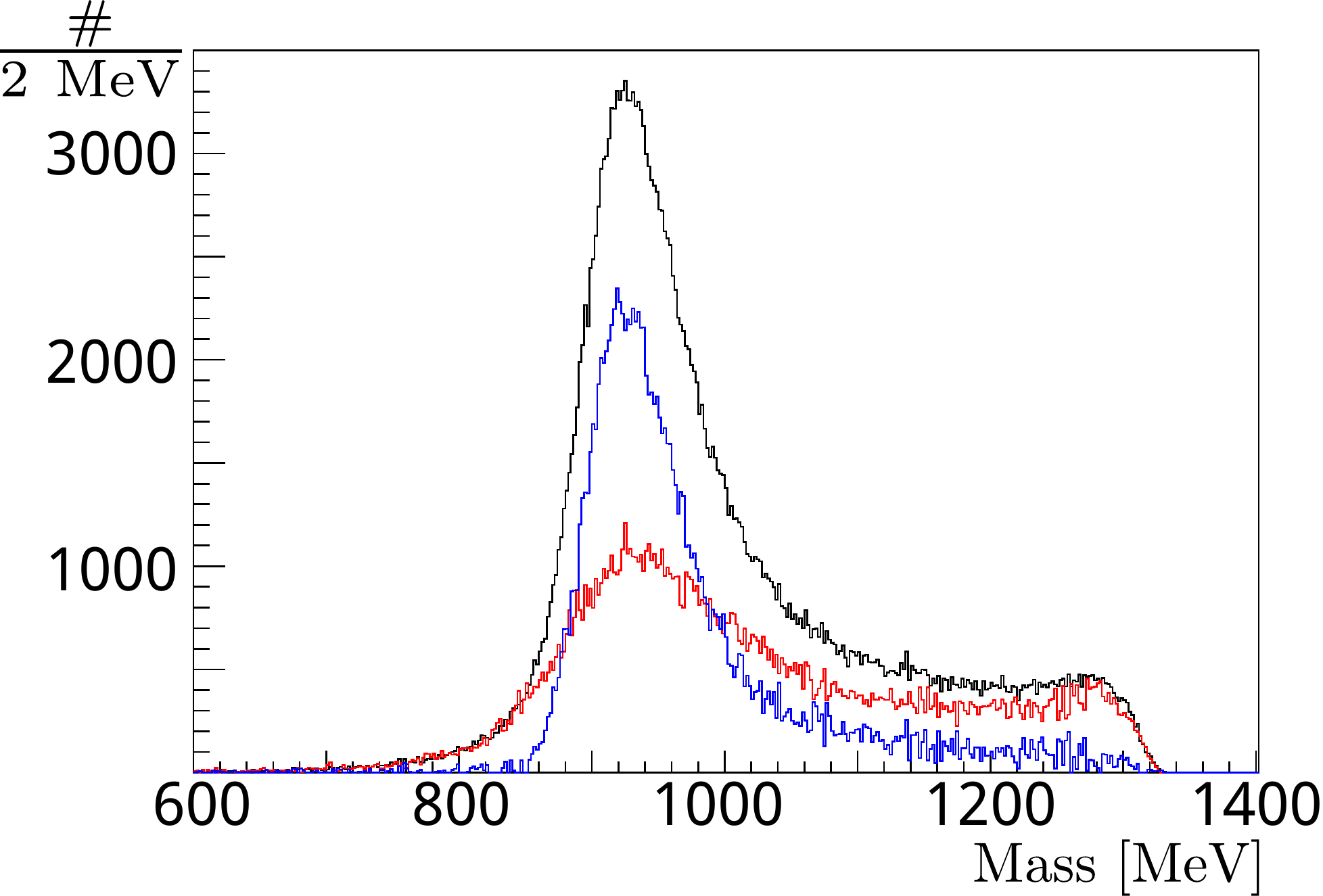}
\caption{\label{pic:missingmass}The missing mass of the system for the three PED-events (top) and the two+three-PED events (bottom) for the butanol data (black), the carbon data (red) and the difference (blue) for a photon energy of $E_\gamma = 1000 \pm 16$~MeV.
}
\end{figure}%
\begin{figure}[tb]
\includegraphics[width=0.4\textwidth]{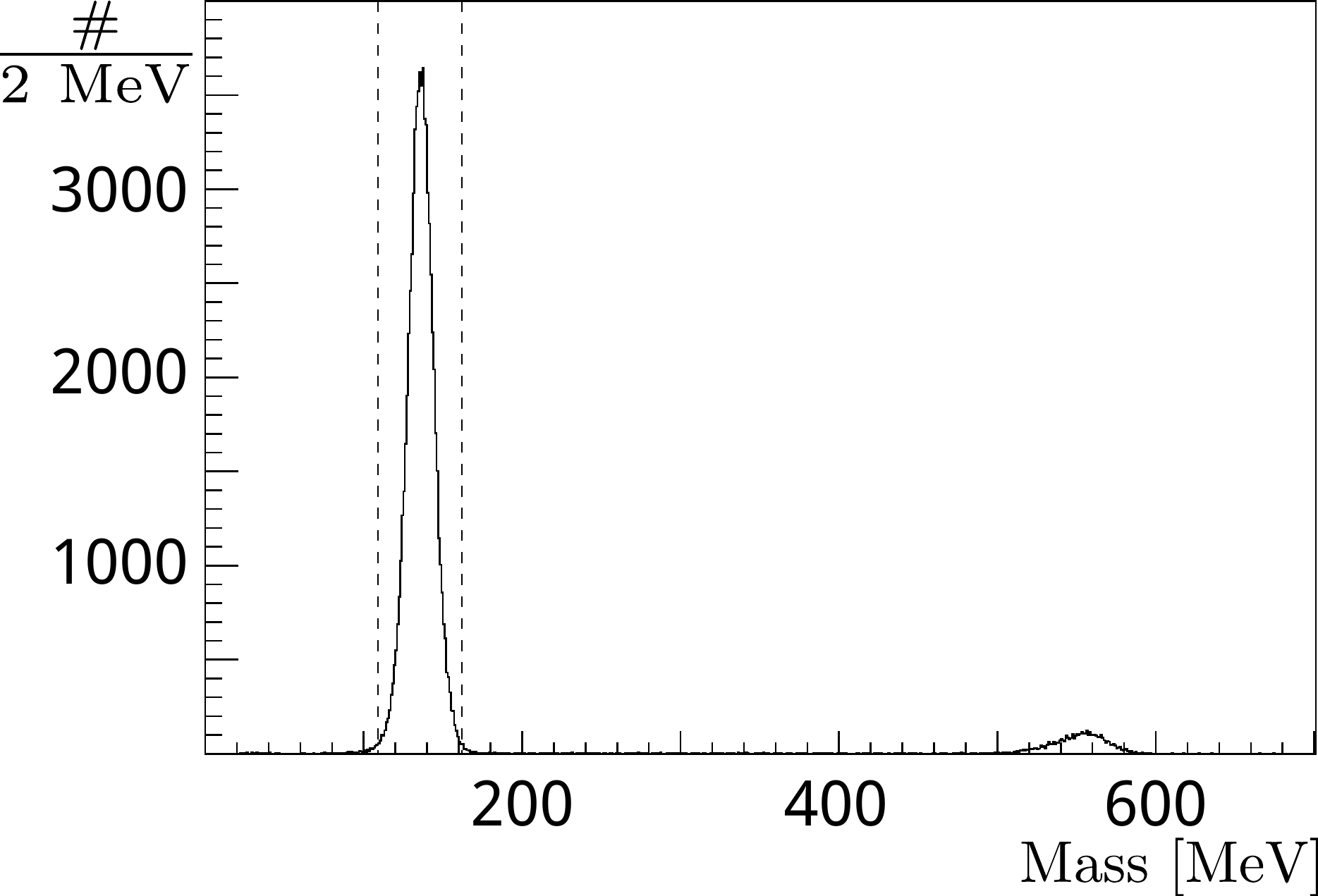}\\[10pt]
\includegraphics[width=0.4\textwidth]{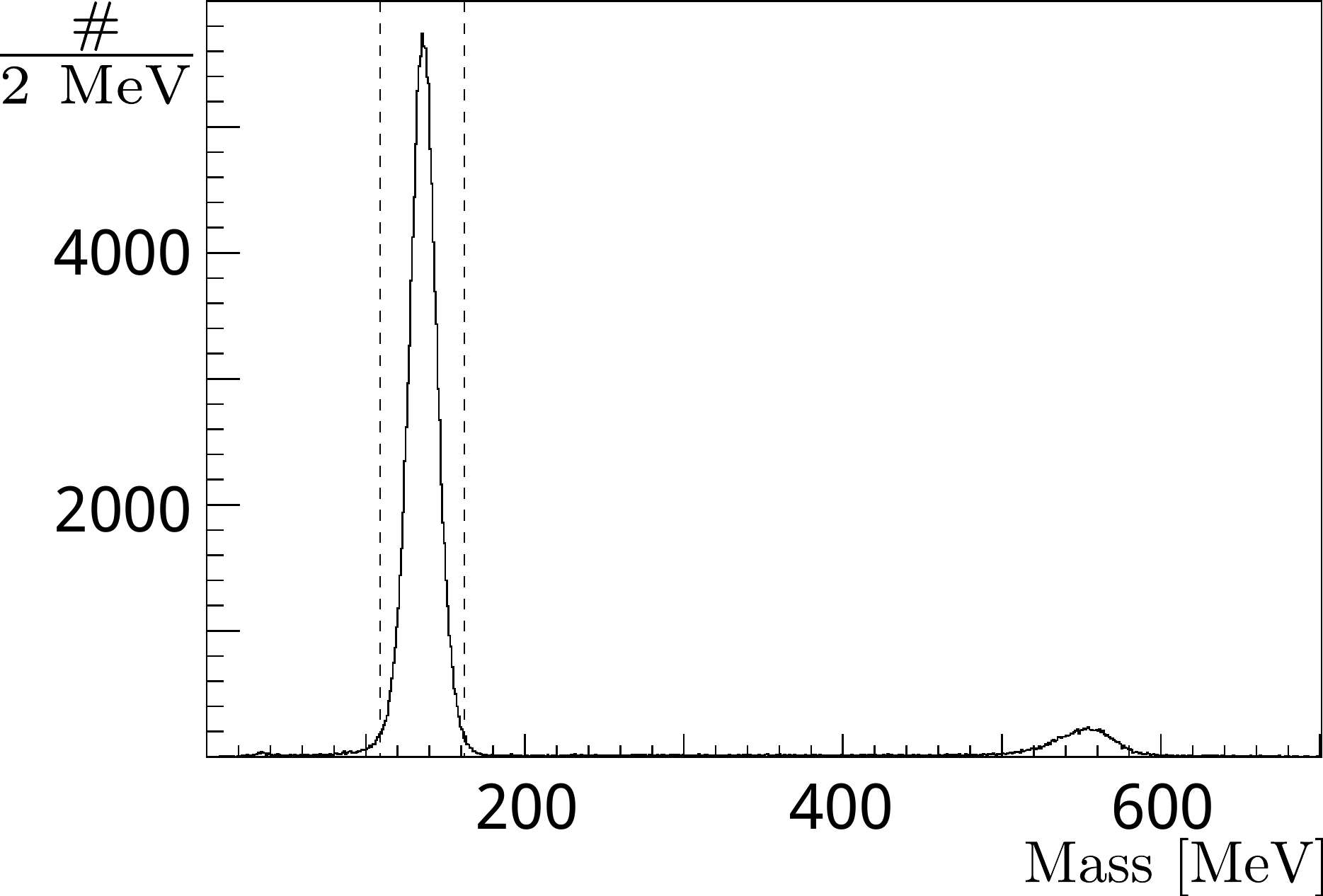}
\caption{\label{pic:invariantmass}The invariant mass of the two photons for the three-PED events (top) and the sum of three and two-PED events (bottom) for the butanol data for a photon energy of $E_\gamma = 1000 \pm 16$~MeV.}
\end{figure}%

For the three-PED events, kinematical cuts were applied to reduce the residual background. The difference in $\phi$ of the proton and pion directions (Fig. \ref{pic:kinematicalcuts}, top) had to be smaller than $|\Delta \phi| < 6^\circ$, and the difference in $\theta$ of the measured and the calculated proton directions had to be smaller than $7^\circ$ (Fig. \ref{pic:kinematicalcuts}, bottom), both representing $\pm1\sigma$ wide cuts. Both distributions are shown for data taken with the carbon foam target and with a butanol target. The difference between the two distributions agrees very well with the distributions that arise when liquid H$_2$ would be used as a target material.

Using the known energy and direction of the beam photon and the momenta of the two photons from the $\pi^0\to\gamma\gamma$ decay, the missing mass was calculated. The missing-mass distribution, shown in Fig. \ref{pic:missingmass} for three and two-PED events, exhibits a clear signal at the mass of the proton. Events satisfying $891< M_p <983$\,MeV, corresponding to a $\pm 1\sigma$ cut, were retained. 

The distribution of the invariant mass of the two decay photons of the selected events is plotted for two (three) PED events in Fig.~\ref{pic:invariantmass}. A clear signal at the pion mass is seen which contains 1.9 (2.7) million events. The background is below 3\%  (0.3\%). At higher mass,  a peak stemming from the $\eta$ meson can be seen around 550~MeV. A $\pm1\sigma$ cut on the invariant mass of the two photons from $m_{\gamma\gamma}=109$\,MeV to $162$\,MeV was used to select the reactions containing a neutral pion in the final state.

Fig. \ref{pic:kinematicalcuts} and \ref{pic:missingmass} show that, due to Fermi motion, the distributions obtained with a carbon target are substantially wider than those obtained when a H$_2$ target is used. The $\pm1\sigma$ kinematical cuts reduce the unwanted contribution of unpolarized nucleons bound in nuclei. For studies of systematic effects, all kinematical cuts were also applied with $\pm3\sigma$ widths. The effect of the cut width on the polarization observables will be discussed in the following chapters.

\section{Extraction of the Polarization Observables}\label{sec:extraction_observables}

\begin{figure}[tb]
\centering
\includegraphics[width=0.4\textwidth]{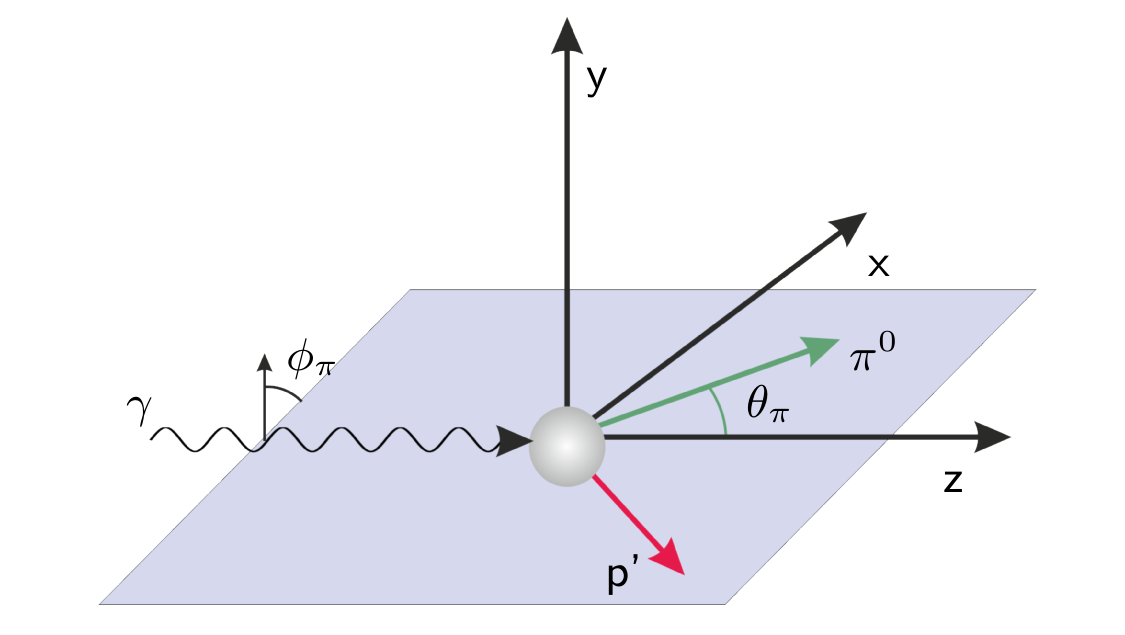}
   \caption{\label{pic:angles}The coordinate definitions in the laboratory system used in this work. The $\phi$ angle is defined between the polarization plane of the photon and the reaction plane, while the $\theta_\pi$ angle describes the angle between the meson and the beam axis. }
\end{figure}

The cross section for photoproduction of pseudoscalar me\-sons using linearly-polarized photons on longitudinally-polarized protons can be written as
\begin{align}
\begin{split}
\frac{d\sigma}{d\Omega}(\theta_\pi,\phi_\pi) = \left.\frac{d\sigma}{d\Omega}\right|_{0} 
\cdot(1 &-p_\gamma \Sigma \cos(2\phi_\pi)\\
&+ p_\gamma p_T G \sin(2\phi_\pi))
\end{split}
\label{eqn:crosssection}
\end{align}
with the unpolarized cross section $\left.\frac{d\sigma}{d\Omega}\right|_{0}$ and the degree of photon ($p_\gamma$) and proton ($p_T$) polarization. The angle definitions can be seen in Fig.~\ref{pic:angles}. Two polarization observables become accessible, the single-polarization observable $\Sigma$, called beam asymmetry, and the double-polarization observable $G$.

The butanol target contains polarizable protons from hydrogen and unpolarizable protons and neutrons bound in carbon and oxygen nuclei. The bound nucleons contribute to the beam asymmetry $\Sigma$ but not to $G$, since they are unpolarizable. The measured distribution of the number of events due to reaction~(\ref{reaction}) with a butanol target $N_{\rm B}$ is given by
\begin{eqnarray}
\hspace{-2mm}\frac{N_{\rm B}(\phi_\pi,\theta_\pi)}{N_{0}(\theta_\pi)} =1-
 p_\gamma \Sigma_{\rm B} \cos(2\phi_\pi) +p_\gamma p_T G_{\rm B}
\sin(2\phi_\pi)
\label{eqn:wq_butanol}
\end{eqnarray}
where $N_0$ is deduced by integrating over $\phi_\pi$. $N_0$ can be decomposed into the contributions of free ($f$) protons $N_{0}^f$ and of nucleons bound ($b$) in nuclei, $N_{0}^b$, with  
 $N_{0}= N_{0}^f + N_{0}^b$. The beam asymmetry $\Sigma_{\rm B}$ and the observable $G_{\rm B}$ off the butanol target are related to the corresponding quantities for scattering off free protons $\Sigma, G$ or bound nucleons by
\begin{eqnarray}
\Sigma_{\rm B} = \frac{N_{0}^f\;\Sigma +
N_{0}^{b}\;\Sigma_{b}}{N_{0}^{f}+N_{0}^{b}};\quad G_{\rm B}
=\frac{N_{0}^f} {N_{0}^{f}+N_{0}^{b}} \cdot G.
\label{eqn:polobs}
\end{eqnarray}
The ratio of the polarizable protons to all protons 
\begin{eqnarray}
d=\frac{N_{0}^f} {N_{0}^{f}+N_{0}^{b}}
\label{eqn:dilution}
\end{eqnarray}
is called the dilution factor and has to be determined to extract the double-polarization observable $G$. The extraction of the dilution factor is explained in section \ref{sec:dilutionfactor} and the extraction of the observable $G$ will be shown in section \ref{sec:g}.

The beam asymmetry $\Sigma$ on the free proton cannot be accessed by the measurement on the butanol target alone, since contributions of bound protons of the carbon nuclei are still present. Therefore further investigations of the influence of bound protons have been carried out, which will be discussed in section \ref{sec:beamasymmetry}.

\begin{figure*}[p]
\centering
\begin{tabular}{cc}
\centering
\resizebox{\textwidth}{!}{%
  \includegraphics{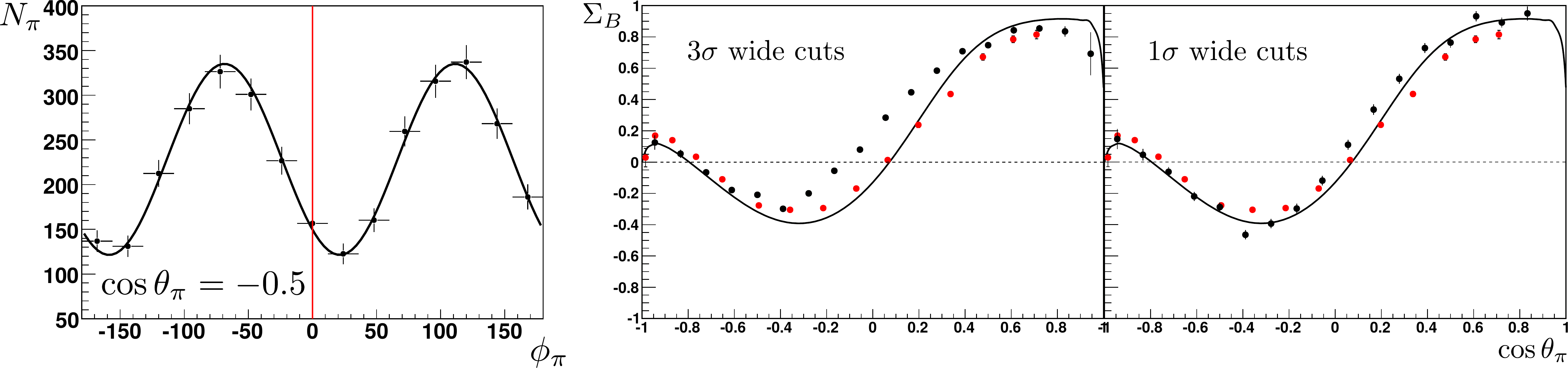}
}
\end{tabular}
\caption{\label{pic:sigma_corrected} Left: An example fit to the $\phi_\pi$ distribution of the pions for $E_\gamma = 800$~MeV. 
Right: The beam asymmetry for $E_\gamma = 1000\pm16$~MeV on the butanol target $\Sigma_B$ (black points) for $3\sigma$ (center) and  $1\sigma$ (right) wide cuts. The data is compared to the GRAAL measurements on free protons (red points) \cite{Bartalini:2005wx}. Only one coherent edge was used here. The curve (a BnGa2014-02 fit) is drawn to guide the eye. \vspace{4mm}}
\end{figure*}

The experimental data have been fitted for each $(E_\gamma,\theta_\pi)$-bin using the function
\begin{eqnarray}
F(\phi_\pi) = A\cdot(1 - B\cos(2\phi_\pi) + C\sin(2\phi_\pi) )
\end{eqnarray}
where the parameters $B$ and $C$ give access to the polarization observables $\Sigma_{\rm B}$ and $G_{\rm B}$. An example fit to the data is shown in Fig. \ref{pic:sigma_corrected}, left.

The double-polarization observable $G$ can also be determined by analyzing the difference of the count rates between two target polarization settings ($\oplus$ and $\ominus$) and the two photon polarization settings ($+45^\circ$ and $-45^\circ$) over the sum:
\begin{eqnarray}
\frac{(N_{\rm B}^\oplus - N_{\rm B}^\ominus)_{+45}-(N_{\rm B}^\oplus - N_{\rm B}^\ominus)_{-45}}{(N_{\rm B}^\oplus + N_{\rm B}^\ominus)_{+45}+(N_{\rm B}^\oplus + N_{\rm B}^\ominus)_{-45}} = p_\gamma p_T \cdot G_{\rm B}\cdot \cos(2\phi_\pi)\nonumber\\
\label{eqn:difference}
\end{eqnarray}
To extract the double-polarization observable $G$, again a correction with the dilution factor is necessary.

\subsection{The Beam Asymmetry $\Sigma$}\label{sec:beamasymmetry}%

According to Eqn.~\ref{eqn:polobs}, the beam asymmetry of the butanol target $\Sigma_{\rm B}$ receives contributions from bound $\Sigma_b$ and free protons $\Sigma$. In a first-order approximation, $\Sigma_b$ could have similar values as the beam asymmetry $\Sigma$ of free protons. Nevertheless, the fraction of events in which $\pi^0$'s are produced off bound protons should be made as small as possible. By using only two-PED events where the proton had been detected, it is also ensured that reactions on neutrons are not contributing.

\begin{figure*}[p]
\centering
\resizebox{0.9\textwidth}{!}{%
  \includegraphics{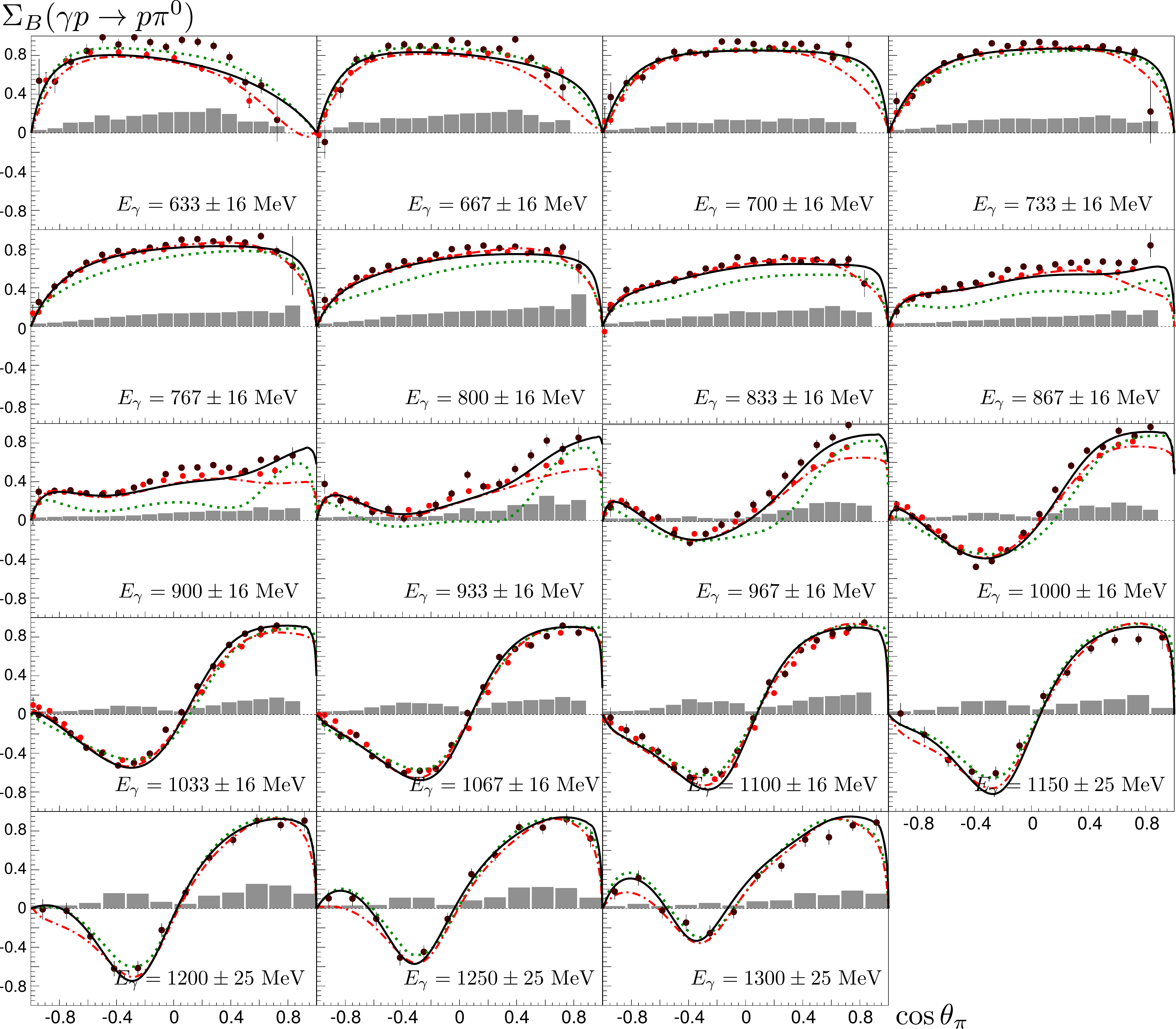}
\vspace{2mm}}
\caption{\label{pic:sigma}The beam asymmetry $\Sigma_{\rm B}$ off the butanol target for different energies (black dots), compared to the GRAAL measurements (red dots) \cite{Bartalini:2005wx} and the PWA solutions: BnGa (black solid line) \cite{Gutz:2014wit}, MAID (green dotted line) \cite{Hilt:2013coa} and SAID (red dashed-dotted line) \cite{Workman:2012jf}.
 \vspace{-2mm}}
\end{figure*}

\begin{figure*}
\centering
\includegraphics[width=0.95\textwidth]{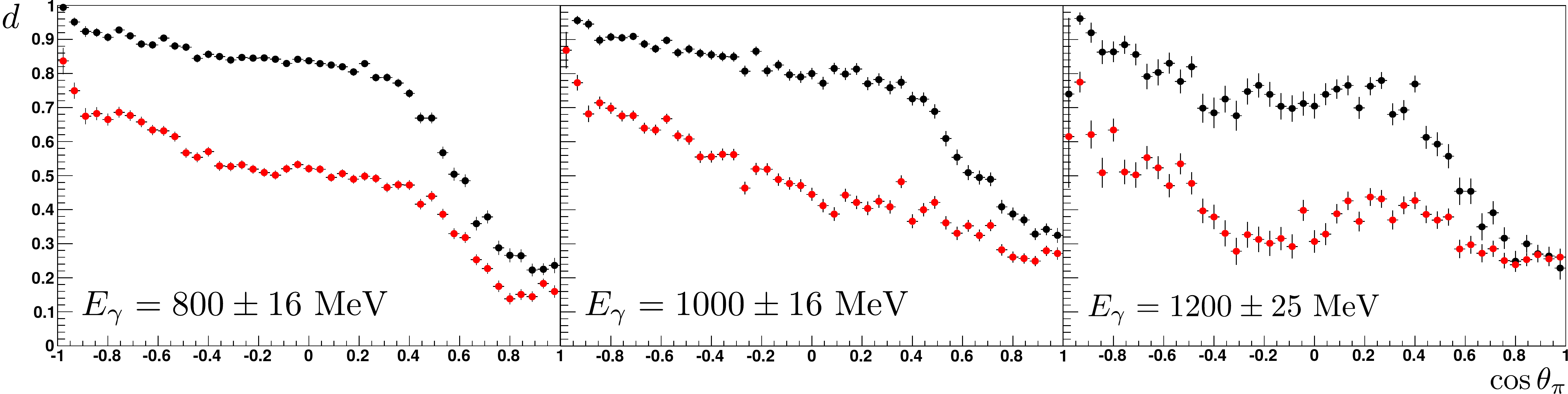}
\caption{\label{pic:dilutionfactor}The dilution factor $d$ with 3$\sigma$ wide cuts (red) compared to the smaller cut widths (black) for three different energies. Only statistical errors are shown. The steep decrease at $\cos \theta_\pi > 0.4$ is due to neutrons contributing in this region.
\vspace{3mm}}
\begin{tabular}{ccc}
\includegraphics[width=\textwidth]{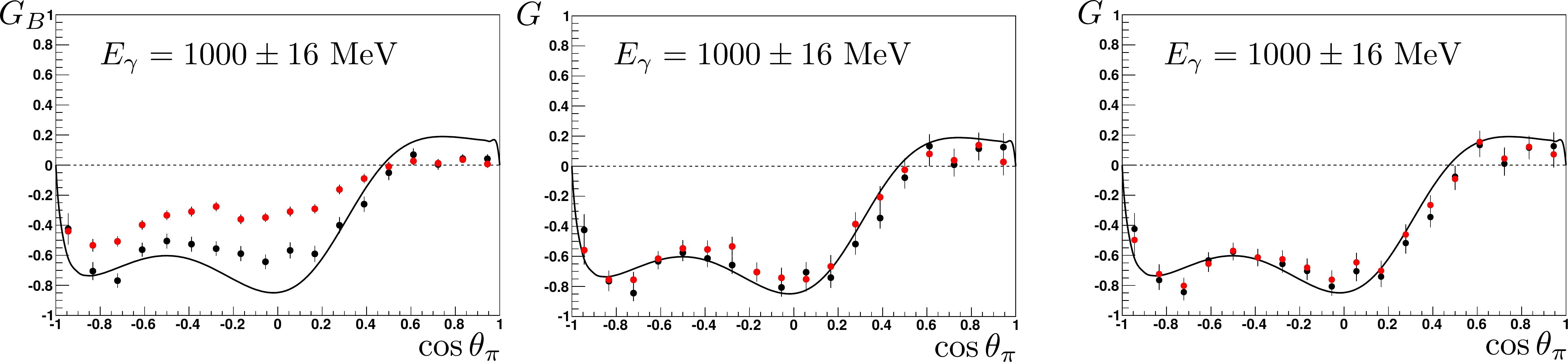}
\end{tabular}
\caption{\label{pic:g_diff_methods}Left: The observable $G_{\rm B}$ without the dilution factor correction for $\pm3\sigma$ (red) and $\pm1\sigma$ (black) wide cuts. Center: $G$ with the dilution factor correction for $\pm3\sigma$ (red) and $\pm1\sigma$ (black) wide cuts. Right: 
The observable $G$ extracted using Eqn.\,\ref{eqn:wq_butanol} (black) or~\ref{eqn:difference} (red). Only statistical errors are shown. The curve (a BnGa fit \cite{Gutz:2014wit}) is drawn to guide the eye.
 }
\end{figure*}

The influence of the cut width of the kinematical cuts ($\Delta \phi$ and $\Delta \theta$) and the missing mass cut on the beam asymmetry $\Sigma_{\rm B}$ can be seen in Fig.~\ref{pic:sigma_corrected}. Applying $\pm3\sigma$ cuts to select reaction~(\ref{reaction}), significant discrepancies are seen between the results using a butanol target and the GRAAL data on free protons \cite{Bartalini:2005wx}. This is due to the larger amount of reactions on carbon nuclei, which are contributing when using wider cuts (cf. Fig \ref{pic:missingmass}). When $\pm1\sigma$ cuts are applied, the beam asymmetry $\Sigma_{\rm B}$ moves significantly closer to the previous GRAAL data on free protons \cite{Bartalini:2005wx}.

The resulting asymmetry $\Sigma_{\rm B}$ for the full energy range from $E_\gamma = 617$~MeV up to $1325$~MeV is shown in Fig. \ref{pic:sigma} and compared to the previous measurements of $\Sigma$ on free protons by the GRAAL collaboration \cite{Bartalini:2005wx}. The good overall consistency demonstrates that the influence of nucleons bound in nuclei can be controlled.  

The systematic errors, shown as a gray band in Fig.~\ref{pic:sigma_corrected}, are deduced from the 
maximal influence of the remaining fraction of carbon contributions in the data on the observable, the differences of the 
different methods outlined above, and the systematic error of the linear photon polarization. The
three components are added quadratically.

\subsection{Determination of the Dilution Factor}\label{sec:dilutionfactor}

The double-polarization observable $G$ receives contributions from the polarized free protons only, see Eqns.~(\ref{eqn:wq_butanol},\ref{eqn:polobs}) or~(\ref{eqn:wq_butanol},\ref{eqn:difference}). The denominator in Eqn.~(\ref{eqn:polobs}) contains the number of events in which a $\pi^0$ is produced off a bound nucleon; hence the fraction of events produced off free protons relative to all produced events, the dilution factor $d$ defined in Eqn.~(\ref{eqn:dilution}), needs to be determined.

$d$ is derived from distributions, which are sensitive on the kinematical properties of the target material. Reactions on bound protons show broader distributions of properly chosen kinematic quantities due to the Fermi motion inside the nucleus.  
In Fig.~\ref{pic:kinematicalcuts} the distribution of the directional differences $\Delta \phi$ -- between the meson and proton momenta -- and $\Delta \theta$ -- between the measured and the calculated proton momenta -- are shown for data taken with a butanol and a carbon foam target. The distribution of the missing mass recoiling against the $\pi^0$ (i.e. the proton) is shown in Fig.~\ref{pic:missingmass}. Again the distributions of the data taken with a carbon target are much broader than those taken with a butanol target.   

The missing mass distributions taken with the butanol target are fitted with the distributions taken with the carbon foam target. The fit range is restricted to the mass region, where reactions without Fermi motion cannot contribute. The fit to the $\phi$ distributions and to the distribution of the missing mass recoiling against the $\pi^0$ results in  dilution factors consistent within 2\%; therefore the latter ones are chosen for further use since these are also defined for the two-PED and not only for three-PED events.

\begin{figure*}
\centering
\resizebox{0.95\textwidth}{!}{%
\includegraphics{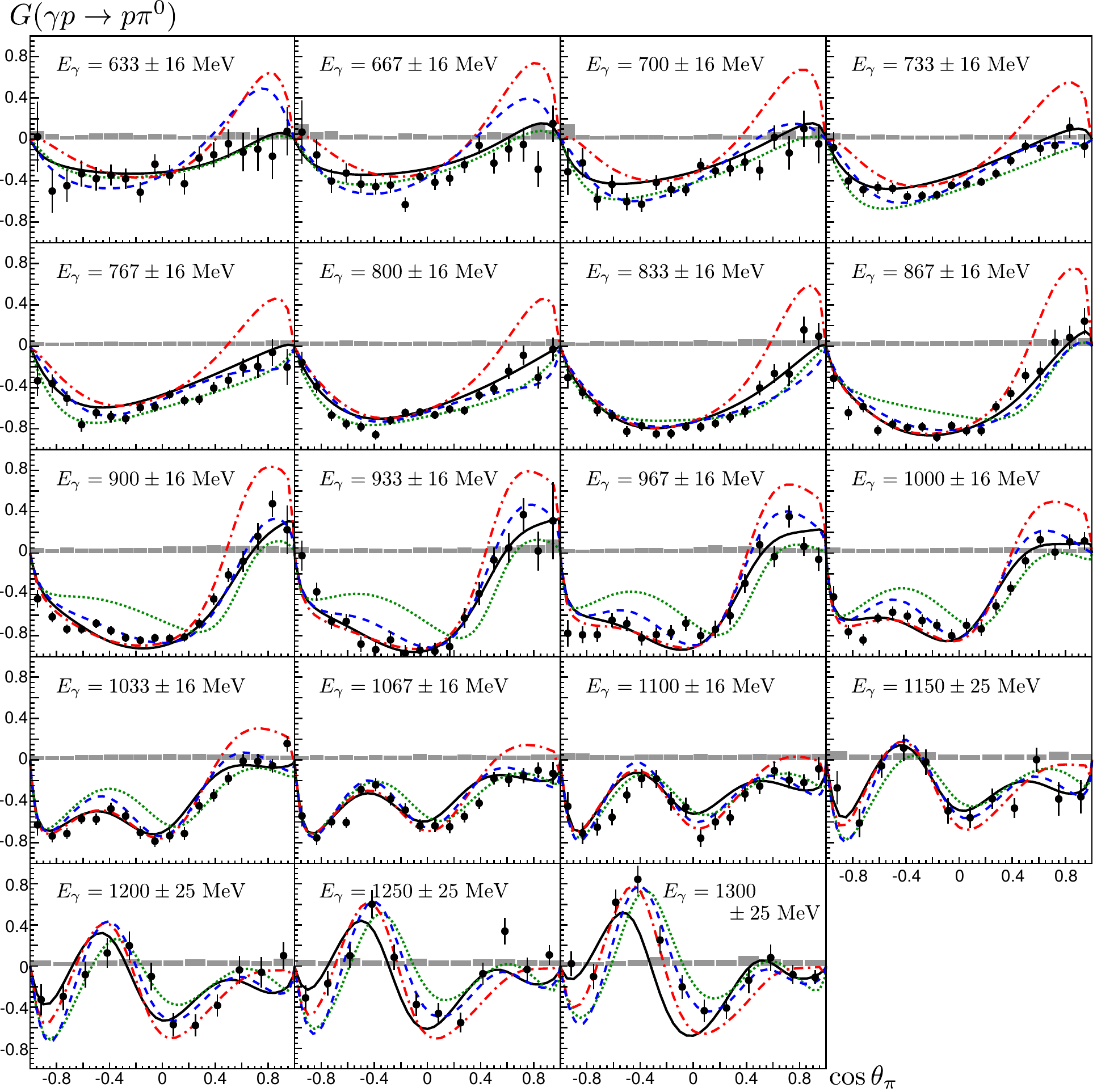}
}
\caption{\label{pic:g_allbins}The double-polarization observable $G$ for all measured photon energies (black dots), compared to the PWA predictions: MAID 2007 \cite{Hilt:2013coa} (green dotted line), SAID CM12 \cite{Workman:2012jf} (red dashed-dotted line), J\"uBo 2013-01 \cite{Ronchen:2014cna} (blue dashed line), and BnGa 2011-02 \cite{Anisovich:2011fc} (black solid line). }
\end{figure*}
\begin{figure*}
\centering
\resizebox{0.95\textwidth}{!}{%
\includegraphics{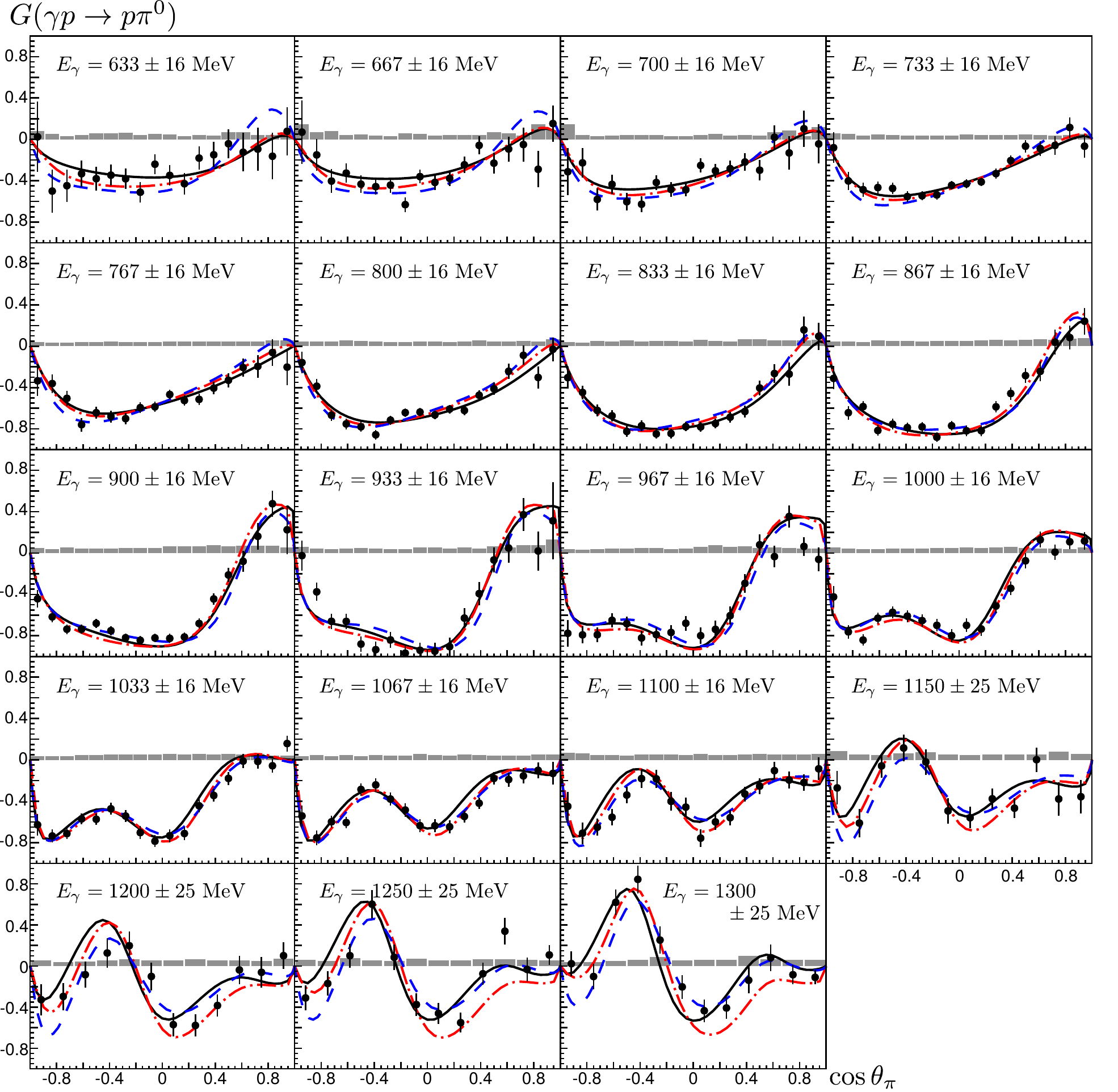}
}
\caption{\label{pic:g_allbins_fits}The double-polarization observable $G$ for all measured photon energies (black dots), compared to the PWA fits to this data set: BnGa 2014-02 \cite{Gutz:2014wit} (black solid line), SAID \cite{Workman:2015} (red dashed-dotted line), and J\"uBo \cite{Roenchen:2015} (blue dashed line).}
\end{figure*}

The dilution factor is a function of the photon energy and of the azimuthal angle of the neutral pion. The results are shown in Fig.~\ref{pic:dilutionfactor} for two different energies.
In the forward direction ($\cos \theta_\pi > 0.4$) a steep decrease of the dilution factor can be observed. This effect is due to the two-PED events where the proton is not detected. Since the charge of the missing nucleon is unknown, reactions on neutrons contribute to the data. This results in a low dilution factor, and the error bars in this $\theta_\pi$ regime increase.

The dilution factor for $\pm3\sigma$ wide kinematical cuts is shown in Fig. \ref{pic:dilutionfactor}, and compared to the  $\pm1\sigma$ cuts. The carbon contributions are much larger with wider cuts, and the dilution factor is smaller over the whole angular range. However, the results on $G$ do not depend significantly on the kinematical cuts (see below).

\subsection{Measurement of the Double-Polarization Observable \boldmath{$G$}}\label{sec:g}%

Figure~\ref{pic:g_diff_methods} shows $G_{\rm B}$ or $G$ as a function of the $\pi^0$ scattering angle for one slice in photon energy in three different plots. The curve represents the BnGa fit~\cite{Gutz:2014wit}. The black points demonstrate the dependence of the result on the method applied.

The left and center panel display the results for two different cut widths -- $1\sigma$ and $3\sigma$ -- without and with dilution-factor correction, which are called $G_{\rm B}$ and $G$, respectively.  Of course, a proper determination of the dilution factor is decisive to obtain reliable results. However, the precise cuts used to determine $G$ have no significant impact on the results: the center panel in Fig.~\ref{pic:g_diff_methods} compares the final result using $\pm1\sigma$ kinematical cuts with results when $\pm3\sigma$ cuts are applied. 

Finally, the double-polarization observable $G$ can be determined in two different ways, as described in equations \ref{eqn:wq_butanol} and \ref{eqn:difference}. In both cases, as shown in Figure~\ref{pic:g_diff_methods} in the right panel, the different methods lead to fully consistent results. The method using Eq. \ref{eqn:wq_butanol} is used to extract the observable, since it exhibits slightly smaller errors.

\begin{figure}[ht]
\resizebox{0.48\textwidth}{!}{%
\includegraphics{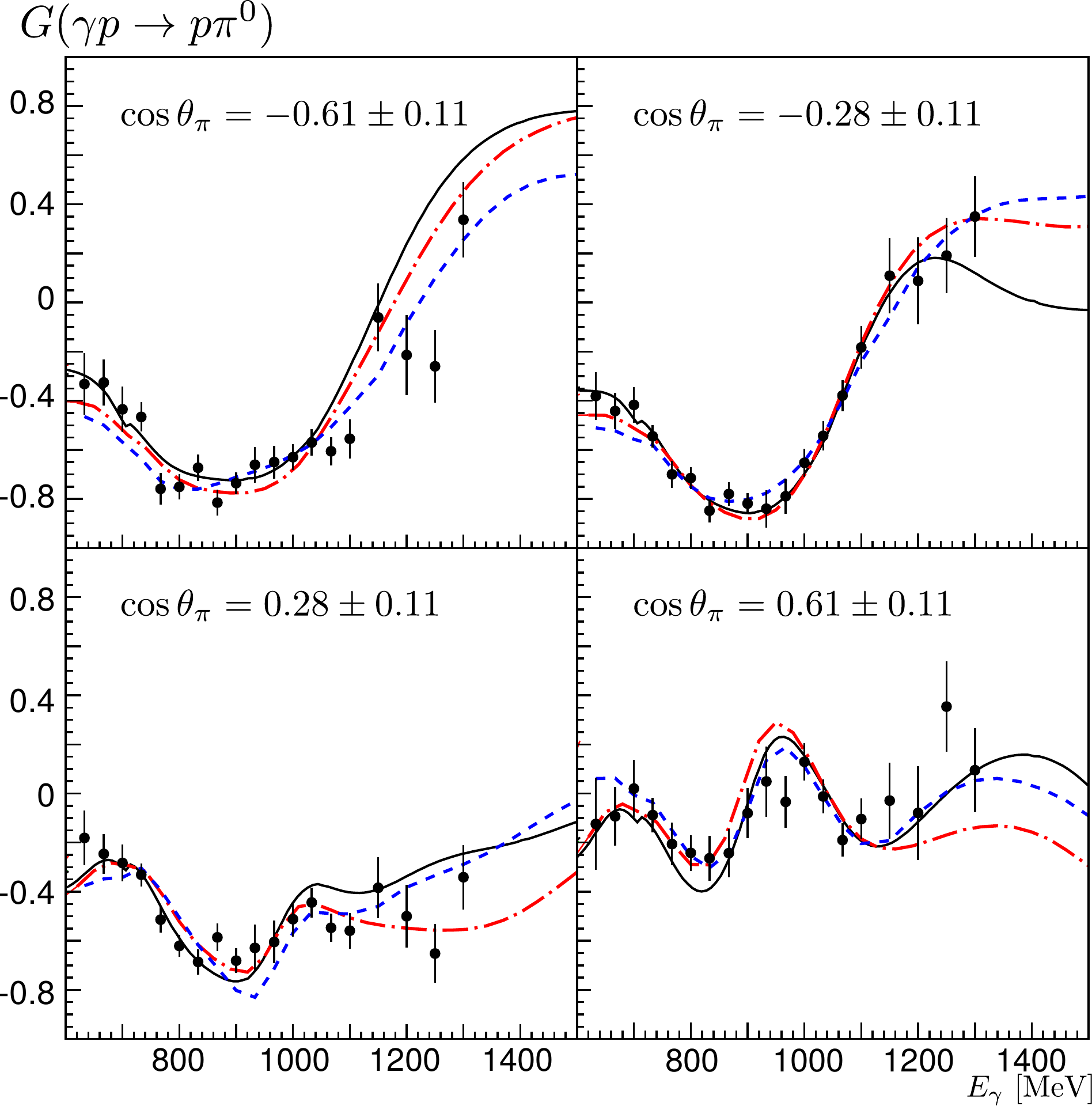}
}
\caption{The energy dependence of $G$ for four different angles, compared to different solutions of the PWAs: BnGa 2014-02 \cite{Gutz:2014wit} (black solid line), SAID \cite{Workman:2015} (red dashed-dotted line), and J\"uBo \cite{Roenchen:2015} (blue dashed line).}
\label{pic:g_diffenergies}    
\end{figure}%
 
Our final results are presented in Fig.~\ref{pic:g_allbins}. The figure shows the double-polarization observable $G$ as a function of $\cos\theta_\pi$ for photon energies from $E_\gamma =617$\,MeV to $E_\gamma =1325$\,MeV. The data and the statistical errors are shown in black, the systematic errors as gray histogram. The systematic errors are derived from five sources:
\begin{enumerate}
\item Variations of the dilution factor when the fit to the missing mass distribution is changed;
\item the uncertainty of the background in the spectrum of invariant masses in the forward direction;
\item the uncertainty of the dilution factor (deduced from a fit to the coplanarity or missing mass distributions);
\item the systematic errors of the photon and the target polarization; 
\item the differences obtained when different methods (Eqn. \ref{eqn:wq_butanol} and \ref{eqn:difference}) are applied.
\end{enumerate}
Symmetry principles enforce $G=0$ for $\cos\theta_\pi=\pm 1$. Data at exactly these points do not exist but the most forward or backward data points are compatible with a curve vanishing at 
$\cos\theta_\pi=\pm 1$. At low photon energies, for $E_\gamma <900$\,MeV, the values of $G$ as functions of $\cos\theta_\pi$ show all negative values, with a single minimum at negative values of $\cos\theta$. 
With increasing photon energy, the distributions become more complicated, with up to three local minima and two local maxima. 
 
In Fig.~\ref{pic:g_allbins}, the data are compared with predictions from different PWA models: with BnGa 2011-02 \cite{Anisovich:2011fc}, MAID 2007 \cite{Hilt:2013coa}, J\"uBo 2013-01 \cite{Ronchen:2014cna}, and SAID CM12 \cite{Workman:2012jf}. The comparison reveals that at lower energies, the MAID and BnGa analyses can describe the data well, while for the SAID results, a deviation at $\cos\theta_\pi \geq 0.4$ becomes apparent. The predictions of J\"uBo show a similar disagreement as SAID for the two lowest photon energy bins at $E_\gamma = 633$ and 667 MeV.
The differences between the predictions of the MAID model and the data in the medium energy region have already been discussed in \cite{thiel:2012} and could be traced back to the multipoles $E_{0+}$ and $E_{2-}$. The largest deviations between the different models can be observed in the higher energy bins ($E_\gamma > 1150$~MeV). These differences most likely occur since resonance contributions in the fourth resonance region are not well known. 

The new $G$ data were communicated to the BnGa, J\"uBo, and the SAID groups and new fits were performed. 
These are presented in Fig.~\ref{pic:g_allbins_fits}. The new fits, BnGa 2014-02 \cite{Gutz:2014wit} (black solid line), SAID \cite{Workman:2015} (red dashed-dotted line), and J\"uBo \cite{Roenchen:2015} (blue dashed line), reproduce the data reasonably well.  Only at higher energies small deviations become visible. For convenience, we show the data in Fig.~\ref{pic:g_diffenergies} for four slices in $\cos\theta$ as a function of the photon energy. All PWA fits can describe the new data very well at lower photon energies, above $E_\gamma > 1150$~MeV the fit results start to diverge. Here more precise data for $G$ are needed to constrain the PWA solutions.

\begin{figure}[pt]
\centering
\includegraphics[width=0.48\textwidth]{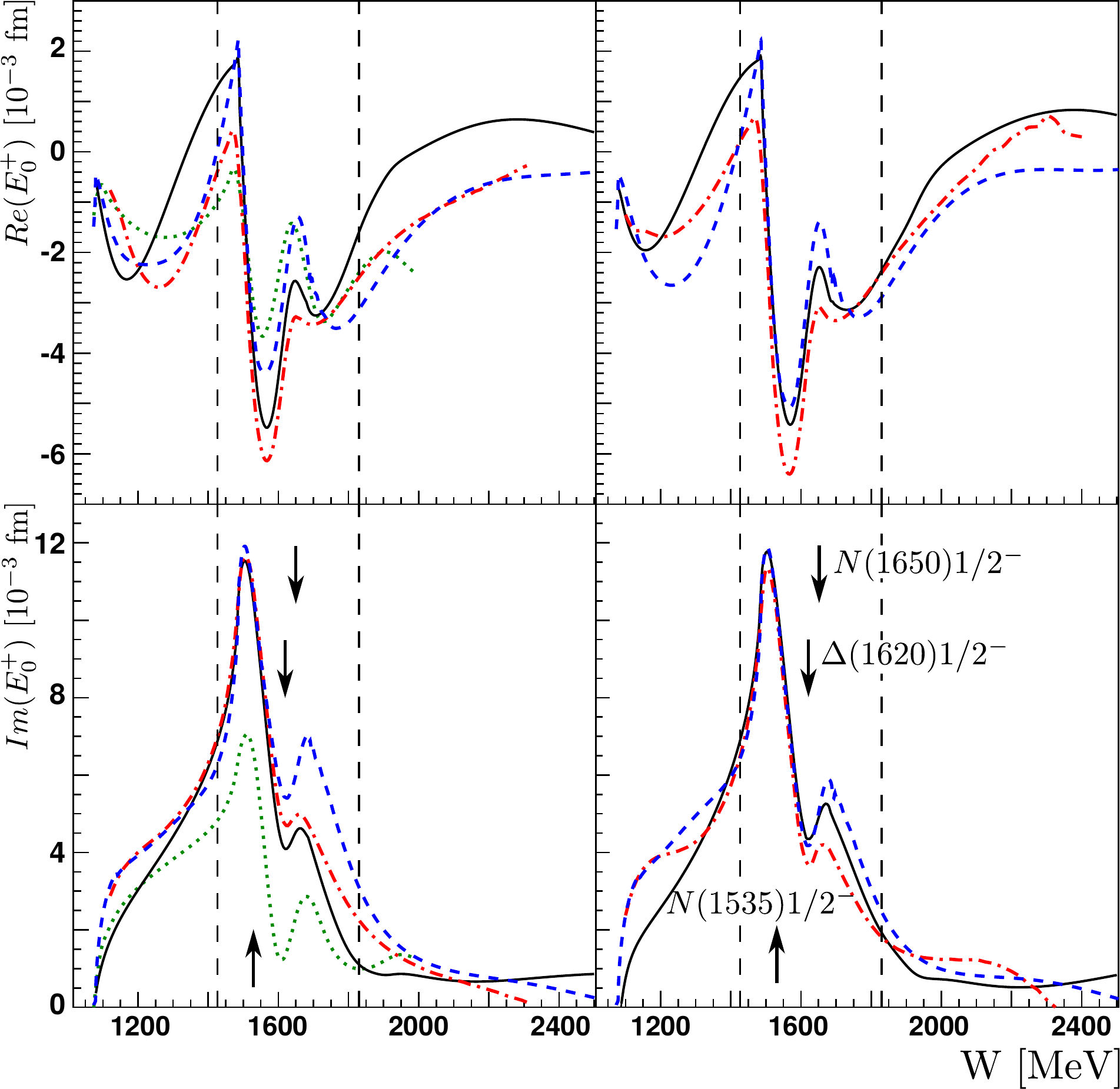}\\
\caption{The real and imaginary part of the multipole $E_{0+}$, before the new data were included (left) and after (right), determined by the BnGa PWA (2011-02 \cite{Anisovich:2011fc} resp. 2014-2 \cite{Gutz:2014wit}: black solid line), the SAID (CM12 \cite{Workman:2012jf} resp. new fit \cite{Workman:2015}: red dashed-dotted line), the J\"uBo model (2015-B \cite{Ronchen:2014cna} resp. new fit \cite{Roenchen:2015}: blue dashed line) and the MAID model (green dotted line) \cite{Hilt:2013coa}. The dashed lines mark the region covered by the observable $G$. Additionally, the positions of the resonances $N(1535)1/2^-$, $N(1650)1/2^-$ and $\Delta(1620)1/2^-$, which contribute to the $E_{0+}$ multipole, are marked with arrows.}
\label{pic:multipoles_m1m}  
\end{figure}

The impact of the new data can be best seen when the multipoles $E_{0+}$, $E_{2-}$ and $M_{2-}$ for the dominantly contributing resonances are compared \cite{Anisovich:2016}. The values of the double-polarization observable $G$ reported here have been used as well as the data on the observables $E$ \cite{Gottschall:2013uha}, $T$, $P$, and $H$ \cite{hartmann:2014mya}. As an example, the real and imaginary parts of the $E_{0+}$ multipole derived from these new fits to the data are shown in Fig. \ref{pic:multipoles_m1m} and compared with the $E_{0+}$ multipole derived from the older fits. For the imaginary part of the multipole, the spread of the solutions reduces considerably: the new data with double-polarization observables ($G, E, T, P, H$) have a decisive influence on the resulting multipoles. For the real part, some reduction in the spread is observed even though much less pronounced. Certainly, more data and further analyses are both required before 
the remaining discrepancies are resolved. 

A longer paper with a comparison of all leading multipoles is in preparation \cite{Anisovich:2016}.

\section{Conclusion}
The first measurement of the double-polarization observable $G$ over a wide angular and photon energy range has been carried out with the CBELSA/TAPS experiment at the ELSA accelerator in Bonn. Simultaneously, the beam asymmetry $\Sigma_{\rm B}$ has been determined using a butanol target and it has been compared to the previous measurements on free protons. Small differences are attributed to bound protons in the carbon atoms of the butanol, as confirmed by a measurement on a pure carbon target.

To extract the observable $G$, a precise determination of the carbon contributions is necessary. With a measurement on a carbon foam target, the energy and angle-dependent dilution factor was extracted. It depends on the kinematic cuts used to extract the observable, hence the dilution factor is not a unique quantity and has to be extracted in correspondence with the observable to be determined.

By comparing the new data on $G$ to the predictions of several partial wave analyses, differences can be observed especially for higher energies. 
The new data sets give new input to the partial wave analyses and allow to constrain the multipole solution of the different analyses. A clear convergence of the leading multipoles is seen, which is the important intermediate step to extract properties of the resonance spectrum of the nucleon. Still more precise measurements of observables in single and double meson production are necessary, in order to reach the final aim: one unique solution for the multipoles and a solution of the mystery of the missing resonances.

\section*{Acknowledgements}
We thank the technical staff of ELSA and the participating institutions for their invaluable contributions to the success of the experiment. We also thank R. Workman, M. D\"oring and D. R\"onchen for providing fits to this new data set and I. Strakovsky and L. Tiator for information about the included data bases.\\
We acknowledge support from the Deutsche Forschungsgemeinschaft (SFB/TR16), the Schweizerischer Nationalfonds, 
the US National Science Foundation, and
the Russian Foundation for Basic Research.



\end{document}